\documentclass{article}
\usepackage{epsfig,subequation,graphics}
\catcode`\@=11

\newcommand{\ov}{{\cal O}}

\newcommand  \f  \varphi

\newcommand{\be}{\begin{equation}}
\newcommand{\ee}{\end{equation}}
\newcommand{\ben}{\begin{displaymath}}
\newcommand{\een}{\end{displaymath}}         
\newcommand{\ba}{\begin{eqnarray}}
\newcommand{\ea}{\end{eqnarray}}
\newcommand{\ban}{\begin{eqnarray*}}
\newcommand{\ean}{\end{eqnarray*}}
\newcommand{\cro}{\dagger}

\newlength{\www}

\def\lesssim{\mathrel{\mathop
  {\hbox{\lower0.5ex\hbox{$\sim$}\kern-0.8em\lower-0.7ex\hbox{$<$}}}}}
\def\greatsim{\mathrel{\mathop
  {\hbox{\lower0.5ex\hbox{$\sim$}\kern-0.8em\lower-0.7ex\hbox{$>$}}}}}
\relax

\begin{document}\vspace{1cm}

\vskip .7cm

\begin{center}
{\Large \bf 
Sudakov Electroweak effects in transversely  polarized beams
}
\vskip .7cm

{\large Paolo Ciafaloni}

{\it INFN - Sezione di Lecce,
\\Via per Arnesano, I-73100 Lecce, Italy
\\ E-mail: Paolo.Ciafaloni@le.infn.it}

\vskip.5cm

{\large Denis Comelli}

{\it INFN - Sezione di Ferrara,
\\Via Paradiso 12, I-35131 Ferrara, Italy\\
E-mail: comelli@fe.infn.it}

\vskip.5cm
{\large Antonio Vergine}

{\it INFN - Sezione di Ferrara,
\\Via Paradiso 12, I-35131 Ferrara, Italy\\
E-mail: vergine@fe.infn.it}

\end{center}

\vskip .5cm

\begin{abstract}
We study Standard Model electroweak radiative corrections for fully
inclusive observables with polarized 
fermionic beams. Our calculations are relevant in view of the possibility
for Next Generation Linear colliders of having 
transversely and/or longitudinally polarized beams.
The case of initial transverse polarization is particularly interesting
because of the interplay of infrared/collinear logarithms 
of different origins, related both to the nonabelian SU(2) and abelian U(1)
sectors.
The Standard model effects turn out to be in the 10\% range at the TeV
scale,
 therefore
particularly relevant in order to disentangle possible New Physics effects.  
\end{abstract}

\setcounter{footnote}{0}
\setcounter{page}{0}

\section{Introduction}

Energy-growing electroweak corrections in the Standard Model have received
recently a lot of 
 attention in the literature, being relevant for LHC physics \cite{acco},
for Next generation of Linear Colliders (NLCs) \cite{nlc}
 and for ultrahigh energy cosmic
rays \cite{berez}. The presence of double logs   
($\log^2\frac{s}{M^2}$ where $M$ is the weak scale)
in one loop electroweak corrections
has been noticed in \cite{first}.
One loop  effects are
typically of the order of 10-20 \%  at the  energy  scale of 1
TeV,  so that the subject of higher orders and/or resummation of large
logarithms has to be addressed. Since two of us \cite{cc} 
made the observation that
double and single logs
that appear in the 1 loop expressions are tied to the infrared structure of
the theory, 
all order resummation has been considered at various levels:
Leading Log (LL) \cite{LL}, Next to Leading Log (NLL) \cite{NLL} and so on.
Moreover many fixed order analyses at the one \cite{oneloop} and two loop
\cite{twoloop} level have been performed.
Although logarithms of infrared-collinear origin also emerge in unbroken
theories like QCD and QED, the presence of symmetry breaking in the
electroweak sector produces large  differences with respect to the 
unbroken gauge theories. For
instance, large double logs of infrared origin are present even in fully
inclusive observables, in contrast to what happens in QED and QCD \cite{ccc}.
 This
in turn implies that the hierarchy of log series is no longer the same as in
QCD, and splitting functions of new kind have to be defined when studying
the analogous of DGLAP equations for EW corrections \cite{cccAP}.

In this paper we consider electro-weak corrections to inclusive observables in the
case of a transverse polarized electron beam. We have in mind in particular
the possibility of having transversely polarized electron beams at NLC
\cite{nlcpol}.
We find that Standard Model
corrections are big enough that one should take them into account when
considering possible New Physics effect in transversely polarized beams
\cite{Rizzo}. 
The analysis of the electroweak IR singularities for the fermionic 
sector in presence of polarized initial states 
is interesting since the final result comes from the interplay of two
distinct effects. On one hand, left fermions are free nonabelian charges
and this produces uncanceled double logs in inclusive quantities; this
phenomenon, related to the SU(2) sector, has been called ``Bloch-Nordsieck
violation'' \cite{ccc}. On the other hand, 
a transversely polarized fermion
is a coherent superposition of  left and right fermions, which have different 
gauge charges; this produces a different but related effect which is
present also in a purely abelian U(1) theory \cite{abelian}.
The interplay between SU(2) and U(1) effects is analyzed in detail in
section 3.

The paper is organized as follows: in next
section we set up the basic formalism for longitudinally and/or
transversely polarized initial beams; this formalism is borrowed largely from
 \cite{transv1,transv2,long}.
In section 3 we analyze the above cross section with
 the formalism of the overlap matrix
 which  give us the rules \cite{ctotal} 
to resum in a straightforward way the double log IR series. 
In section 4 we analyze the phenomenological implications 
related  to initial beams polarization in the
process  $e^+ e^-\to q \bar{q}+X$ at high energy.

\section{Cross section with polarized initial beams}
We consider the inclusive 
process $e^+ e^-\to q \bar{q}+X$, where $X$ includes  $\gamma,Z,W^\pm$
radiation
and where we sum over final quark flavors
 (all the fermions are supposed massless). The initial electron momentum is
along the positive $z$ axis. The
initial beam is transversely polarized, the initial electron 
being polarized along the positive $x$ axis and the positron in the
opposite direction. $\theta$ and $\phi$ are the usual polar and azimuthal
angles of the outgoing quark momentum  in this frame.
 To describe the initial spin
 states in $e^{+}\,e^{-}$ we choose the helicity states as basis which is 
convenient at high energies.
We indicate the helicity
amplitudes with $M^{f_1f_2}_{i_1i_2}$, where $i_1(i_2)$ is the helicity of 
the initial electron (positron), while
$f_1(f_2)$  corresponds to the final   quark (antiquark).
\begin{figure}
      \centering
      \includegraphics[height=90mm]
                  {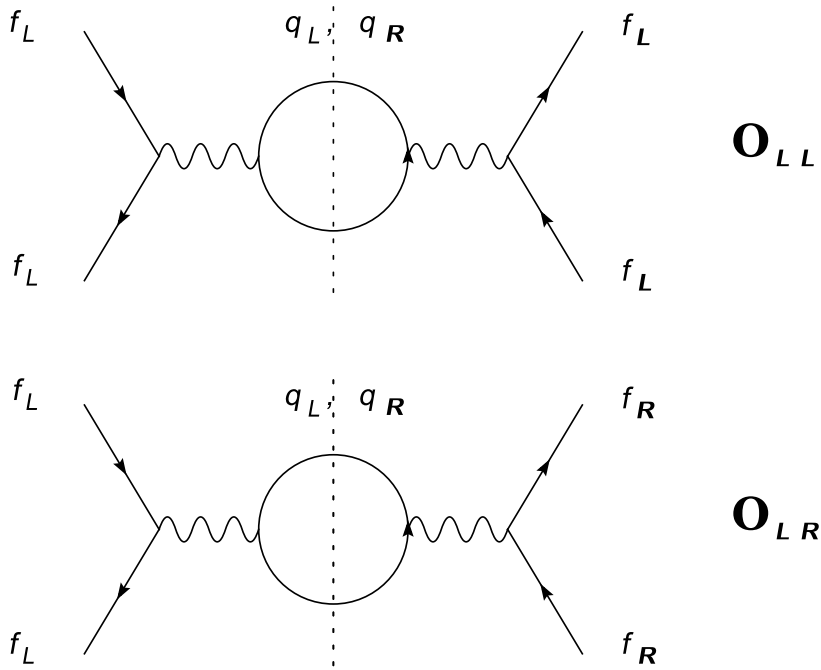}
 \caption{\label{overlap} 
The overlap matrix for initial L fermions and in the mixed L-R case,
relevant for transverse polarization.
%Diagrammatic picture of the  contributions 
%to Overlap function
}
     \end{figure}

Let us now consider the process $e^+e^-\to q\bar{q}+X$. We denote the
longitudinal and transverse components of the $e^-(e^+)$ polarizations by 
$P_L(\bar{P}_L)$ and $P_T(\bar{P}_T)$ and we assume that the two transverse
polarization vectors are parallel up to a sign\footnote{for a pure initial
  state $|P_L|^2+|P_T|^2=1$, while for a partially polarized beam
 $|P_L|^2+|P_T|^2<1$; see \cite{transv1}}. For instance for a
``natural'' transverse polarization, where the electron and positron
polarizations are opposite, we have $P_T=1,\bar{P}_T=-1$.
For unpolarized beams we have $P_{L}=\bar{P}_L=P_{T}=\bar{P}_{T}=0$.
Finally, if both beams are polarized left, we have $P_{L}=\bar{P}_L=1$.
Then we obtain 
the averaged squared amplitude\cite{transv1,Rizzo}:
\be\bar{|M|^2}=\label{dpewq}
\sum_{f_1f_2}\left\{
\frac{(1-P_L)(1-\bar{P}_{ L})}{4}
|M^{f_1f_2}_{+-}|^2+\frac{(1+P_L)(1+ \bar{P}_{ L} )}{4}|M^{f_1f_2}_{-+}|^2
+\frac{P_T\;  \bar{P}_{ T}}{2}
Re\{e^{2i\phi}M^{f_1f_2}_{+-}M^{f_1f_2*}_{-+}\} \right\}
\ee

\section{Overlap formulation}

Let us now rewrite eqn. (\ref{dpewq}) in the language of the overlap
matrix, $\hat {\ov}={\cal S}{\cal S}^\cro$, ${\cal S}$ being the
  S-matrix. We define $\hat{\ov}$ (see fig. \ref{overlap}) 
as an operator with four isospin indices (see \cite{ccc} for details). All the
final phase space factors that are  relevant for cross section calculations
are included, so that:
\ba\nonumber
\hat \ov_{i_1i_2}^{j_1j_2}(s,\theta,\phi)=
\sum_{t,Y}
 O^{(t,Y)}(\theta,\phi) \;\hat{{\cal P}}^{(t)\,j_1j_2}_{i_1i_2}\qquad
\ea
Here $\hat{{\cal P}}^{(t)}$ are 
4 indices operators in isospin space, whose expressions
 are given in
our previous works\cite{ctotal} and \cite{abelian}, 
while $ O^{(t,Y)}(\theta,\phi)$ are the
coefficients  of the overlap with
defined total t-channel isospin $t$ and total t-channel hypercharge $Y$.
We can then rewrite eqn. (\ref{dpewq}) in the form
\be
\frac{d\sigma}{d\phi d\cos\theta}=\frac{N_cN_f}{256\pi^2s}\left\{
(1-P_{L})(1-\bar{P}_L)O_{RR}^{(0,0)}+
(1+P_{L})(1+\bar{P}_L)
(O_{LL}^{(0,0)}\pm O_{LL}^{(1,0)})
+(1\pm 1) P_T \bar{P}_{T}O_{LR}^{(\frac{1}{2},y_L^e-y_R^e)}
\cos2\phi\right\}
\ee
where  $N_f$ is the number of families, $N_c$ the number of colors
and the + (-) sign is relative to $\sigma_{e^-e^+}$ ($\sigma_{e^-\bar{\nu}}$)
where we have explicitly written the cross section dependence on the
initial beam polarization.
%factorized out the spin density structure of the 
%cross section.
\begin{figure}
%\begin{minipage}[b]{0.5\linewidth}
      \centering
      \includegraphics[height=80mm]{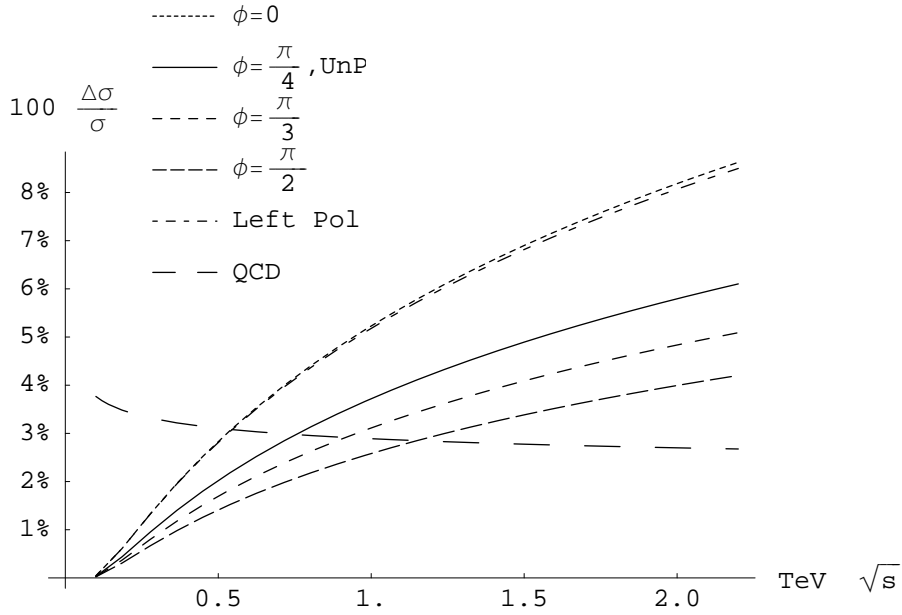}
 \caption{\label{theta} Resummed double log corrections 
to the $\theta$-integrated cross
 section for transversely polarized beams and for various values of $\phi$
as a function of the c.m. energy; the $\phi=\pi/4$ case correspond to the
unpolarized cross section case, while the QCD line is
 relative to the one loop $\alpha_S(\sqrt{s})/\pi$ 
contribution. }
%\end{minipage}
\end{figure}
By inserting the Standard Model weak couplings, we find the following
tree level values of the various coefficients:
\begin{subequations}\begin{eqalignno}
O_{LL}^{(t=0,Y=0)}&=\left(
\frac{3}{16}g^4+\frac{1}{2}g'^4 y_{iL}^2\sum_f y_{fL}^2\right)
(1+\cos\theta)^2+\frac{1}{2}\; g'^4 y_{iL}^2\sum_f y_{fR}^2 (1-\cos\theta)^2
\\
O_{LL}^{(t=1,Y=0)}&=\left(-\frac{1}{16}g^4+\frac{1}{2}g'^4 
y_{iL}^2\sum_f y_{fL}^2\right)(1+\cos\theta)^2
+\frac{1}{2}g'^4 y_{iL}^2\sum_f y_{fR}^2(1-\cos\theta)^2
\\
O_{LR}^{(t=\frac{1}{2},Y=y^e_L-y^e_R)}&= g'^4
  y_{iL}\;y_{iR}\sum_f(y_{fR}^2+y_{fL}^2)\;
{\sin^2\theta}
\\
O_{RR}^{(t=0,Y=0)}&= g'^4\;y_{iR}^2\sum_f y_{fR}^2(1+\cos\theta)^2
+g'^4\; y_{iR}^2\sum_f y_{fL}^2(1-\cos\theta)^2
\end{eqalignno}\end{subequations}

The evolution in double log approximation is given by the weights 
 $\;e^{-{\bf L}_W(t(t+1)+\tan^2\theta_W \;Y^2)}\;$
where $\;{\bf L}_W=\frac{g^2}{16 \pi^2}\log^2\frac{s}{M_W^2}$ and where 
$t$ and $Y$ are respectively the total isospin and the total
 hypercharge in the t channel. The effect of all orders resummation is
 therefore given by the following substitutions \cite{ctotal}:
\be
O^{(t,Y)}(\theta,\phi)\to e^{-{\bf L}_W
\left(t(t+1)+
\tan^2\theta_W Y^2 \right)}\; \; O^{(t,Y)}(\theta,\phi)
\ee
so that the full improved leading double logs $e^+e^-$ cross section
is:
\ba\label{vacca}
\frac{d\sigma_{e^+e^- }}{d\phi d\cos\theta}&=&\frac{N_cN_f}{256\pi^2s}\left\{
(1-P_{L})(1-\bar{P}_L)O_{RR}^{(0,0)}+
(1+P_{L})(1+\bar{P}_L)
(O_{LL}^{(0,0)}+
e^{-2{\bf L}_W}O_{LL}^{(1,0)})\right.
\\
&+& \left.  2 P_T \bar{P}_{ T}e^{-{\bf L}_W(\frac{3}{4}+
\tan^2\theta_W (y_L^e-y_R^e)^2)}O_{LR}^{(\frac{1}{2},y_L^e-y_R^e)}
\cos2\phi \right \} \nonumber
\ea
This expression features an interesting effect in the mixed L-R channel
$O_{LR}$
where both SU(2)($\sim \frac{3}{4}{\bf L}_W$) and U(1)
($\sim{\bf L}_W \tan^2\theta_W (y_L^e-y_R^e)^2$) 
related effects are present. In fact this channel is characterized both by
a nonzero t-channel isospin ($T=\frac{1}{2}$) and a nonzero hypercharge
$y_L^e-y_R^e$. 
Overall, the analysis of the electroweak mass singularities for the fermionic 
sector in presence of polarized initial states turns out to be  quite 
interesting, not only for the presence 
of IR singularities generated by the existence of free non abelian 
asymptotic states, but also for the fact that
the  asymptotic states can be a coherent superposition of different
gauge eigenstates: in this case, a transverse polarized fermion
is a coherent superposition of  left and right fermionic 
gauge charges.
Such a mixing phenomenon is quite common in the EW theory: 
in the bosonic sector the Z and the photon are a superposition of
the $B_{\mu}$ and the $W^3_{\mu}$ gauge fields, giving rise to a peculiar
pattern of EW double log corrections \cite{ccclong}.
 In the scalar sector the  higgs and the longitudinals goldstone bosons 
are a superposition of  the gauge doublets $\Phi$ and 
$\Phi^*$ (having opposite U(1) charges), and Block-Nordsieck violation is
present even in abelian theories \cite{abelian}.

\begin{figure}
%\begin{minipage}[b]{0.5\linewidth}
       \centering
\includegraphics[height=80mm] {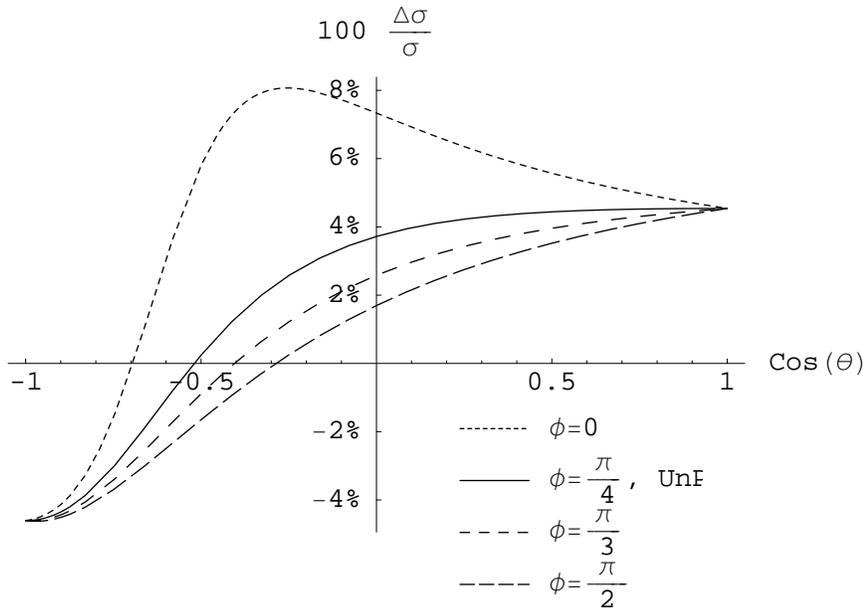}
 \caption{\label{phi} Resummed double log corrections 
to the differential cross section  for transversely polarized beams
at $\sqrt{s}$= 1 TeV for various values of $\phi$.}
%\end{minipage}
     \end{figure}
\begin{figure}[t]
      \centering
      \includegraphics[height=80mm]
                  {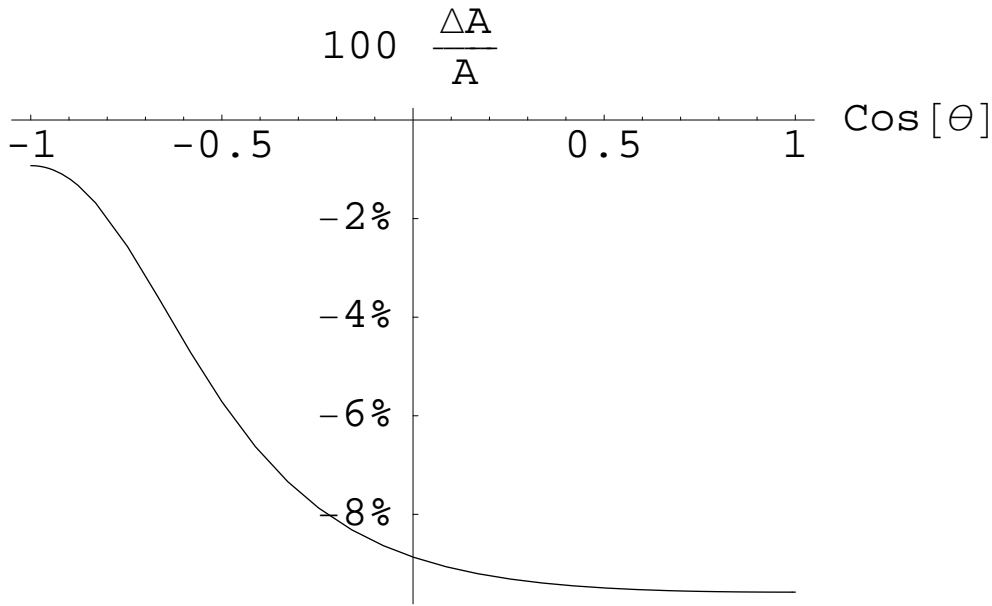}
 \caption{\label{asimm}EW corrections as a function of the $\theta$ angle
 to the Asymmetry of eq.\ref{ass} when the beam energy is 1 TeV. }
     \end{figure}

In figs. \ref{theta},\ref{phi},\ref{asimm} various combinations of the
resummed differential cross section  (\ref{vacca}) are plotted
in the case of   transversely polarized beams
($ P_{L}= \bar{P}_{L}=0,P_{T}=1,\bar{P}_T=1$).
Namely, in fig. 2  the percent corrections 
for the integrated cross section
 $\int^1_{-1}  d\cos \theta\frac{d\sigma}{d\phi d\cos\theta}$ 
are plotted at different values of the $\phi$ angle as function of the
c.m. energy.   In fig.3 corrections to the 
the differential cross section 
$\frac{d\sigma}{d\phi d\cos\theta} $ at one TeV
 is plotted for different values of the azimuthal  angle.

An interesting observable that singles out the transverse $\phi$ dependent
piece of the cross section is the asymmetry
\be\label{ass}
A(\cos \theta,s)=\frac{\int^{\pi/4}_0 d\phi\;\sigma\;- 
\int^{\pi/2}_{\pi/4} d\phi\;\sigma\; }{\int^{\pi/2}_{0} d\phi\;\sigma\; }
\ee
which results particularly sensitive to the EW IR corrections. A plot of
this quantity is given in fig. \ref{asimm}.

Overall, we see that EW double log resummed corrections for transversely
polarized initial beams are well into the 10 \% range, so that this kind of
corrections have to be taken into account when considering possible New
Physics effects \cite{Rizzo}.

The full approach to such a kind of phenomena requires the introduction of 
``anomalous'' structure functions which doesn't have 
a clear  probabilistic interpretation.
As it is easy to see in fig.1, the ``diagonal'' overlap matrix
$\hat{\ov}_{LL}$ can be 
rephrased in terms of the convolution of structure functions for
left  handed fermions, 
the ``mixed'' one $\hat{\ov}_{LR}$ 
 requires the introduction of a third king of mixed
structure function with two different legs (left and right).
Clearly the naive classical probabilistic interpretation
fails to explain such a structure being 
generated by the
quantum phenomenon of the coherent superposition of
the wave functions.
The full set of evolution eqs of the SM
 taking into account such a phenomena will be given soon in \cite{2c}.

\vspace{1.cm}
{\rm \bf Acknowledgements}: We are grateful to Marcello Ciafaloni
 for discussions and  participation to the first part 
of our  results.

\begin{thebibliography}{99}

\bibitem{acco} E. Accomando, A. Denner and  S. Pozzorini,
Phys. Rev. D {\bf 65} 073003 (2002);
E. Maina, S. Moretti, M.R. Nolten and  D.A. Ross,
Phys. Lett. B {\bf 570} 205 (2003). 

\bibitem{nlc}TESLA: The Superconducting Electron Positron Linear Collider
  with an integrated X-ray laser laboratory. Technical Design Report. Part
  3. Physics at an e+ e- linear collider.

\bibitem{berez}
V. Berezinsky, M. Kachelriess, 
S. Ostapchenko Phys.Rev.Lett. {\bf 89} 171802 (2002);
C. Barbot, M. Drees, Astropart.\ Phys.\  {\bf 20}, 5 (2003).

\bibitem{first}
M.~Kuroda, G.~Moultaka and D.~Schildknecht,
%``Direct One Loop Renormalization Of SU(2)-L X U(1)-Y Four Fermion Processes And Running Coupling Constants,''
Nucl.\ Phys.\ B {\bf 350} (1991) 25;
G.~Degrassi and A.~Sirlin,
%``Gauge invariant selfenergies and vertex parts of the Standard Model in the pinch technique framework,''
Phys.\ Rev.\ D {\bf 46}, 3104 (1992);
A.~Denner, S.~Dittmaier and R.~Schuster,
%``Radiative corrections to gamma gamma $\to$ W+ W- in the electroweak standard model,''
Nucl.\ Phys.\ B {\bf 452}, 80 (1995);
A.~Denner, S.~Dittmaier and T.~Hahn,
%``Radiative corrections to Z Z $\to$ Z Z in the electroweak standard model,''
Phys.\ Rev.\ D {\bf 56}, 117 (1997),
%A.~Denner and T.~Hahn,
%``Radiative corrections to W+ W- $\to$ W+ W- in the electroweak standard  model,''
Nucl.\ Phys.\ B {\bf 525}, 27 (1998);
W.~Beenakker, A.~Denner, S.~Dittmaier, R.~Mertig and T.~Sack,
%``High-energy approximation for on-shell W pair production,''
Nucl.\ Phys.\ B {\bf 410}, 245 (1993);
W.~Beenakker, A.~Denner, S.~Dittmaier and R.~Mertig,
%``On shell W pair production in the TeV range,''
Phys.\ Lett.\ B {\bf 317}, 622 (1993).


\bibitem{cc}P.~Ciafaloni, D.~Comelli,
%``Sudakov enhancement of electroweak corrections,''
Phys.\ Lett.\ B {\bf 446}, 278 (1999).


\bibitem{LL}
V.~S.~Fadin, L.~N.~Lipatov, A.~D.~Martin and M.~Melles,
%``Resummation of double logarithms in electroweak high energy processes,''
Phys.\ Rev.\ D {\bf 61} (2000) 094002;
P.~Ciafaloni, D.~Comelli,
%``Electroweak Sudakov form factors and nonfactorizable soft QED effects  at NLC energies,''
Phys.\ Lett.\ B {\bf 476} (2000) 49.
M.~Melles,
%``Subleading Sudakov logarithms in electroweak high energy processes to  all orders,''
Phys.\ Rev.\ D {\bf 63}, 034003 (2001);
Phys.\ Rev.\ D {\bf 64}, 014011 (2001);
Phys.\ Rev.\ D {\bf 64}, 054003 (2001);
Eur.\ Phys.\ J.\ C {\bf 24}, 193 (2002);
Phys.\ Rept.\  {\bf 375}, 219 (2003); 
\bibitem{NLL}
J.~H.~Kuhn, A.~A.~Penin and V.~A.~Smirnov,
%``Summing up subleading Sudakov logarithms,''
Eur.\ Phys.\ J.\ C {\bf 17}, 97 (2000);
J.~H.~Kuhn, S.~Moch, A.~A.~Penin, V.~A.~Smirnov,
%``Next-to-next-to-leading logarithms in four-fermion electroweak  processes at high energy,''
Nucl.\ Phys.\ B {\bf 616}, 286 (2001)
[Erratum-ibid.\ B {\bf 648}, 455 (2003)]

\bibitem{oneloop}
M.~Beccaria, P.~Ciafaloni, D.~Comelli, F.~M.~Renard, C.~Verzegnassi,
%``Logarithmic expansion of electroweak corrections to four-fermion  processes in the TeV region,''
Phys.\ Rev.\ D {\bf 61} (2000) 073005;
%M.~Beccaria, P.~Ciafaloni, D.~Comelli, F.~M.~Renard and C.~Verzegnassi,
%``The role of the top mass in b production at future lepton colliders,''
Phys.\ Rev.\ D {\bf 61} (2000) 011301;
M.~Beccaria, F.~M.~Renard, C.~Verzegnassi,
%``Logarithmic SUSY electroweak effects on four-fermion processes at TeV  energies,''
Phys.\ Rev.\ D {\bf 63} (2001) 095010;
%M.~Beccaria, F.~M.~Renard and C.~Verzegnassi,
%``Top quark production at future lepton colliders in the asymptotic  regime,''
Phys.\ Rev.\ D {\bf 63} (2001) 053013;
%M.~Beccaria, F.~M.~Renard and C.~Verzegnassi,
%``The role of universal and non universal Sudakov logarithms in four
%fermion processes at TeV energies: The one-loop approximation revisited,''
Phys.\ Rev.\ D {\bf 64}, 073008 (2001); 
Nucl.\ Phys.\ B {\bf 663}, 394 (2003);
M.~Beccaria, S.~Prelovsek, F.~M.~Renard, C.~Verzegnassi,
%``Top quark production at TeV energies as a potential SUSY detector,''
Phys.\ Rev.\ D {\bf 64}, 053016 (2001);
M.~Beccaria, M.~Melles, F.~M.~Renard, C.~Verzegnassi,
%``SUSY scalar production in the electroweak Sudakov regime of lepton  colliders,''
Phys.\ Rev.\ D {\bf 65} (2002) 093007;
M.~Beccaria, M.~Melles, F.~M.~Renard, S.~Trimarchi, C.~Verzegnassi,
%``Sudakov expansions at one loop and beyond for charged scalar and  fermion pair production in SUSY models at future linear colliders,''
Int.\ J.\ Mod.\ Phys.\ A {\bf 18}, 5069 (2003);
M.~Beccaria, F.~M.~Renard, S.~Trimarchi, C.~Verzegnassi,
%``Charged Higgs production in the 1-TeV domain as a probe of  supersymmetric models,''
Phys.\ Rev.\ D {\bf 68} (2003) 035014;
%M.~Beccaria, F.~M.~Renard and C.~Verzegnassi,
%``Reliability of a high energy one-loop expansion of e+ e- $\to$ W+ W- in  the SM and in the MSSM,''
%Nucl.\ Phys.\ B {\bf 663}, 394 (2003);
A.~Denner and S.~Pozzorini,
%``One-loop leading logarithms in electroweak radiative corrections. I:  Results,''
Eur.\ Phys.\ J.\ C {\bf 18}, 461 (2001) and  
Eur. Phys. J. C {\bf 21}, 63 (2001).

\bibitem{twoloop}
~Hori, H.~Kawamura, J.~Kodaira,
%``Electroweak Sudakov at two loop level,''
Phys.\ Lett.\ B {\bf 491} (2000) 275;
W.~Beenakker, A.~Werthenbach,
%``New insights into the perturbative structure of electroweak Sudakov  logarithms: Breakdown of conventional exponentiation,''
Phys.\ Lett.\ B {\bf 489}, 148 (2000);
%W.~Beenakker and A.~Werthenbach,
%``Electroweak two-loop Sudakov logarithms for on-shell fermions and  bosons,''
Nucl.\ Phys.\ B {\bf 630} (2002) 3;
A.~Denner, M.~Melles, S.~Pozzorini,
%``Two-loop electroweak angular-dependent logarithms at high energies,''
Nucl.\ Phys.\ B {\bf 662}, 299 (2003);
U.~Aglietti, R.~Bonciani,
%``Master integrals with one massive propagator for the two-loop  electroweak form factor,''
Nucl.\ Phys.\ B {\bf 668}, 3 (2003).
See also S.~Pozzorini,
%``Electroweak Radiative Corrections At High Energies,''
arXiv:hep-ph/0201077 and references therein
%%CITATION = HEP-PH 0201077;%%;

\bibitem{ccc}M.~Ciafaloni, P.~Ciafaloni, D.~Comelli,
%``Bloch-Nordsieck violating electroweak corrections to inclusive TeV  scale hard processes,''
Phys.\ Rev.\ Lett.\  {\bf 84}, 4810 (2000) and 
%``Electroweak Bloch-Nordsieck violation at the TeV scale: 'Strong' weak  interactions?,''
Nucl.\ Phys.\ B {\bf 589} (2000) 359

\bibitem{cccAP}
M.~Ciafaloni, P.~Ciafaloni, D.~Comelli,
%``Towards collinear evolution equations in electroweak theory,''
Phys.\ Rev.\ Lett.\  {\bf 88}, 102001 (2002)
\bibitem{nlcpol}
G.~Moortgat-Pick, H.~M.~Steiner,
%``Physics opportunities with polarized e- and e+ beams at TESLA,''
Eur.\ Phys.\ J.\ directC {\bf 3}, 6 (2001);
S.~Hesselbach, O.~Kittel, G.~Moortgat-Pick, W.~Oeller,
%``New Ideas on SUSY Searches at Future Linear Colliders,''
arXiv:hep-ph/0310367.

\bibitem{Rizzo}T.~G.~Rizzo,
%``More transverse polarization signatures of extra dimensions at linear  colliders,''
JHEP {\bf 0308}, 051 (2003) and 
JHEP {\bf 0302}, 008 (2003);
B.~Ananthanarayan, S.~D.~Rindani,
%``CP violation at a linear collider with transverse polarization,''
arXiv:hep-ph/0309260.

\bibitem{transv1}
K. Hikasa, Phys.Rev.{\bf D33}, 3203 (1986); Phys.Rev.{\bf D38}, 1439 (1988)

\bibitem{transv2}
F.M. Renard Z.Phys.{\bf C45}:75,1989;
J. Fleischer, K. Kolodziej, F. Jegerlehner Phys.Rev. {\bf D49} 2187 (1994) 

\bibitem{long}
B.W. Lynn, C. Verzegnassi
Phys.Rev.{\bf D35} 3326 (1987)
 
\bibitem{ctotal}
M.~Ciafaloni, P.~Ciafaloni, D.~Comelli,
Phys.\ Lett.\ B {\bf 501}, 216 (2001)

\bibitem{abelian}
M.~Ciafaloni, P.~Ciafaloni, D.~Comelli,
%``Bloch-Nordsieck violation in spontaneously broken Abelian theories,''
Phys.\ Rev.\ Lett.\  {\bf 87}, 211802 (2001)

\bibitem{ccclong}
M.~Ciafaloni, P.~Ciafaloni, D.~Comelli
Nucl.\ Phys.\ B {\bf 613}, 382 (2001)

\bibitem{2c}
P.~Ciafaloni and D.~Comelli, to appear.
\end{thebibliography}
\end{document}